\documentclass[aps,prd,showpacs,twocolumn]{revtex4}
\usepackage[usenames,dvipsnames]{color}
\usepackage{epsfig}

\def\QW{Q_W(^{133}_{\phantom{k} 55}\text{Cs}) }

\begin{document}

\title{Constraining Extra Neutral Gauge Bosons with Atomic Parity Violation Measurements}
\author{Ross Diener$^{1,2}$\footnote{Email: rdiener@perimeterinstitute.ca}, 
Stephen Godfrey$^{3}\footnote{Email: godfrey@physics.carleton.ca}$ 
and Ismail Turan$^3$\footnote{Email: ituran@physics.carleton.ca}}
\affiliation{
$^1$ Department of Physics \& Astronomy, McMaster University, Hamilton, ON, Canada, L8S 4M1\\
$^2$ Perimeter Institute for Theoretical Physics, Waterloo, ON, Canada, N2L 2Y5\\
$^3$Ottawa-Carleton Institute for Physics, Department of Physics, Carleton University, Ottawa, ON, Canada, K1S 5B6\\
}

\date{\today}

\begin{abstract}

The discovery of a new neutral gauge boson, $Z^{\prime}$, could provide the first
concrete evidence of physics beyond the standard model. We explore how future parity violation experiments, especially atomic parity violation (APV) experiments, can be used to constrain $Z^{\prime}$ bosons. 
We use the recent measurement of the $^{133}$Cs nuclear weak charge to estimate lower bounds 
on the mass of $Z^{\prime}$ bosons for a number of representative models and to put
constraints on the couplings of a newly discovered $Z'$ boson.  
We also consider how these constraints might be improved by future APV experiments that will 
measure nuclear weak charges of multiple isotopes. We show how measurements 
of a single isotope, and combining measurements into ratios and differences, 
can be used to constrain the couplings of a $Z'$ and discriminate between models.  We include
in our results the constraints that can be obtained from the
experiments {\tt Qweak} and P2  that measure the proton weak charge.
We find that current and future parity violation experiments could potentially play an important 
role in unravelling new physics if a $Z'$ were discovered.

\end{abstract}
\pacs{14.70.Pw,12.60.-i,12.15.Mm} 

\maketitle

\section{Introduction}

The Large Hadron Collider (LHC) has started to explore the TeV energy regime
opening up the possibility of discovering new fundamental particles. One such particle,
which arises in many models of physics beyond the Standard Model (SM) and should be
relatively straightforward to discover, is a new, massive, spin-1, s-channel resonance 
($Z^{\prime}$)~\cite{Hewett:1988xc,Langacker:2008yv,Rizzo:2006nw,Leike:1998wr,Cvetic:1995zs,Godfrey:1994qk,Diener:2010sy,Erler:2011ud}. 
Although such a resonance could arise as a Kaluza-Klein excitation of the photon or the 
SM $Z$ \cite{ArkaniHamed:1998rs,Randall:1999ee,Davoudiasl:2000wi}, 
we restrict our attention to $Z^{\prime}$s arising from an extended gauge
symmetry. For these models,
contributions to precision electroweak observables 
generally imply a mass bound $M_{Z^{\prime}} \gtrsim 1$ 
TeV~\cite{Chivukula:2002ry, Erler:2009jh,arXiv:1005.3998}, 
and recent estimates~\cite{Diener:2010sy,Erler:2011ud} indicate that the LHC should be able 
to probe far beyond these bounds, up to $\sim$~5 TeV once the LHC reaches its design 
energy and luminosity.

While direct detection of new particles is unambiguous,
precision measurements provide a complementary 
approach to exploring new physics \cite{RamseyMusolf:1999qk,Erler:2004cx,Erler:2011sv}. 
For example, precision electroweak (EW) measurements impose 
comparable or stronger 
bounds than direct detection on many $Z^{\prime}$ bosons 
\cite{Chivukula:2002ry, Erler:2009jh,arXiv:1005.3998}, 
and EW
constraints on oblique parameters~\cite{peskin} are difficult to avoid in other new 
physics scenarios. One important recent update to the EW observables is a better than 1\% extraction 
of the $^{133}$Cs weak charge \cite{Porsev:2010de} from APV
experiments \cite{wieman}
\begin{equation} \label{result}
 \QW = -73.16(29)_{\text{exp}}(20)_{\text{th}}.
\end{equation}
The new determination is in perfect agreement with the SM prediction of
$\QW = -73.16(3)$ \cite{Nakamura:2010zzi}. It has since been noted that this measurement provides 
strong constraints on the $S$ parameter~\cite{Hobbs:2010fu} and it can act as the dominant observable 
in global analysis
of precision measurements used to constrain models of new physics~\cite{Hsieh:2010zr}.
Thus, $\QW$ has been measured to a precision that can significantly constrain new physics
\cite{RamseyMusolf:1999qk}.

In this paper we examine the implications of APV experiments on the physics
of $Z^{\prime}$ bosons. We consider both the $\QW$ result and observables from a number of future 
APV experiments which are expected to perform measurements of weak charges along the isotope chains 
of Ba~\cite{barium}, Fr~\cite{francium, triumf}, Ra~\cite{radium} and Yb~\cite{ytterbium}.  
For completeness, we also consider the implications of other experiments measuring parity violation;
the {\tt Qweak} measurement of the proton weak 
charge at Jefferson Lab~\cite{Gericke:2010zz}, the P2 experiment at Mainz \cite{p2@mami}, 
and the SoLID deep inelastic measurement at Jefferson Lab \cite{Souder:2011zz,solid,Mantry:2010ki}.
The presence of a $Z^{\prime}$ would result in an 
${\cal O}(M_Z^2 g_{Z^{\prime}}^2 / M_{Z^{\prime}}^2 g_Z^2)$ correction to a weak charge.
This effect has been discussed in the literature~\cite{marciano,Langacker:1990jf, COLO-HEP-228-REV, RamseyMusolf:1999qk, hep-ph/0410260},
but only a 
small subset of $Z^{\prime}$ models were considered. We expand on these earlier studies
by considering 
$Z^{\prime}$s from the following classes of models: little Higgs (LH) 
\cite{ArkaniHamed:2002qy,Schmaltz:2004de,Kaplan:2003uc}, 
left-right symmetric (LR), 
technicolor (TC) \cite{Chivukula:1994mn, Chivukula:1995gu, Simmons:1996ws,Hill:1994hp, Lane:1995gw, 
Lane:1996ua, Lane:1998qi, Popovic:1998vb}, 
3-3-1 \cite{Pisano:1991ee}, 
E$_6$ \cite{Hewett:1988xc}, and the ununified model (UUM) \cite{Georgi:1989xz}. 
Refs.~\cite{Hewett:1988xc,Langacker:2008yv,Rizzo:2006nw,Leike:1998wr,Cvetic:1995zs,Godfrey:1994qk, 
Diener:2010sy} contain details of these models and their phenomenology. 

Our analysis consists of two parts.  We first examine how measurements in APV experiments, including
the $\QW$ result,  
can be used to bound the $Z^{\prime}$ mass for various models.  
Given the sensitivity of APV measurements to $Z'$
physics, we also examine how they could be used to constrain the properties of a newly discovered $Z'$.
We find that APV experiments can provide strong constraints on the
$u$- and $d$-quark couplings of a new $Z^{\prime}$ that are otherwise difficult to obtain, and 
could provide valuable information that complements other measurements.

The paper is organized as follows:  in Section \ref{sec:apv} we outline the 
equations and conventions we use in our analysis. 
In the remaining subsections, we discuss mass bounds coming from $\QW$, weak charge ratios 
in isotopes of Fr and Yb, and the proton weak charge, and other parity violation experiments. In Section \ref{sec:coupl} we examine 
how these measurements
can constrain the couplings of a $Z^{\prime}$ and help distinguish between different models. 
We summarize our results in Section \ref{sec:conc}.

\section{New Gauge Bosons and Weak Charges}\label{sec:apv}

The nuclear weak charge of an isotope of element $X$ 
with $Z$ protons and $N$ neutrons, or $A=N+Z$ nucleons, can be written \cite{RamseyMusolf:1999qk}
\begin{equation}
 Q_W(^{A}_{Z}X) =  Q_W^0 (^{A}_{Z}X)+  \Delta Q_W(^{A}_{Z}X)
\end{equation}
where a 0 superscript will denote a SM prediction, and a $\Delta$ will denote a new physics contribution. For the nuclear weak charge, the tree level SM prediction is given by 
\cite{Nakamura:2010zzi} 
\begin{eqnarray}
  Q_W^0(^{A}_{Z}X) &=& -4\ c_A^e \left[ (2Z+N)c_V^u + (Z+2N)c_V^d  \right] \nonumber\\
  &=& -N + Z \left( 1 - 4 s_W^2 \right)
\end{eqnarray}
where $c_{A,V}^f\equiv c_L^f\mp c_R^f$ are the SM $Z$ boson couplings to fermions  
and $s_W^2\equiv \sin^2\theta_W$, with $\theta_W$ being the weak mixing angle~\cite{Nakamura:2010zzi}. 

At the sub-MeV energies of atomic physics, we can use the effective Lagrangian to describe the SM neutral current interaction between an electron $e$ and 
a fermion $f$~\cite{Nakamura:2010zzi,RamseyMusolf:1999qk}. It has the following relevant parity-violating terms~\cite{RM-comment}:
\begin{eqnarray} \nonumber \label{SM}
 {\cal L}_{\text{PV}}^f &=& - \frac{g_{Z}^2}{4M_{Z}^2} \left( c_A^e \bar{e}
\gamma_\mu \gamma_5 e \right) ( c_V^f \bar{f} \gamma^\mu f )\\
 &\equiv&\ \   \frac{g_{Z}^2}{16M_{Z}^2} Q_W^f ( \bar{e}
\gamma_\mu \gamma_5 e ) ( \bar{f} \gamma^\mu f )
\end{eqnarray}
where  $g_Z\equiv g_2/\cos\theta_W$ and $g_2$ is the gauge 
coupling constant 
of $SU(2)_L$. It is understood that ${\cal L}_{\text{PV}}=\sum_f {\cal L}_{\text{PV}}^f$.
Similarly, we can write the effective Lagrangian for the neutral current interaction of a $Z^\prime$ boson with mass $M_{Z^\prime}$
\begin{eqnarray} \label{effLag} \nonumber
\!\!\!\!\!\!\Delta {\cal L}_{\text{PV}}^f\!\! &=& - \frac{g_{Z^\prime}^2}{4M_{Z^\prime}^2} ( \tilde{c}_A^e \bar{e}
\gamma_\mu \gamma_5 e )
(  \tilde{c}_V^f \bar{f} \gamma^\mu f )\\
&=&   \frac{g_{Z}^2}{16M_{Z}^2}  \left[-4 \frac{M_{Z}^2}{M_{Z^\prime}^2} 
\frac{g_{Z^\prime}^2}{g_{Z}^2} \tilde{c}_A^e \tilde{c}_V^f
\right] ( \bar{e}
\gamma_\mu \gamma_5 e )(  \bar{f} \gamma^\mu f )
\end{eqnarray}
where $\tilde{c}_{A,V}^f$ and $g_{Z^{\prime}}$  are defined for the $\bar{f}fZ^\prime$ interactions 
in analogy with the SM quantities in Eqn.~\ref{SM}. 

It is straightforward to obtain the new physics contribution to the weak charge of a particle
from Eqn.~\ref{effLag}, and for the proton and neutron, we find
\begin{eqnarray} \label{np} \nonumber
 \Delta Q_W^p &=& -4 \frac{M_Z^2}{M_{Z^{\prime}}^2} \frac{g_{Z^{\prime}}^2}{g_Z^2} \tilde{c}_A^e (2 \tilde{c}_V^u + \tilde{c}_V^d)\;, \\ 
 \Delta Q_W^n &=& -4 \frac{M_Z^2}{M_{Z^{\prime}}^2} \frac{g_{Z^{\prime}}^2}{g_Z^2} \tilde{c}_A^e (2 \tilde{c}_V^d + \tilde{c}_V^u)\;.
\end{eqnarray}

The corrections to the weak charge of a given isotope can now be built from the above quantities
\begin{equation} \label{nucl}
\Delta Q_W(^{A}_{Z} X)  =  Z \Delta Q_W^p + N \Delta Q_W^n\;.
\end{equation}
From Eqns. \ref{np} and \ref{nucl} it can be seen that a new neutral gauge boson does indeed modify 
weak charges by contributions of order
${\cal O}(M_Z^2 g_{Z^{\prime}}^2 / M_{Z^{\prime}}^2 g_Z^2)$.

\subsection{The Weak Charge of $^{133}$Cs}

Using Eqn.~\ref{nucl}, a precise determination of any weak charge can be readily translated into a 
bound on $Z^{\prime}$ physics.
For example, the recent measurement of $\QW$ can be used to obtain a mass bound on a model with 
fixed $Z^\prime$ couplings.
We find, as in Ref.~\cite{Porsev:2010de},  that the $Z^\prime$ arising in the E$_6$ $\chi$ 
model would result in a correction of
$\Delta Q_W(^{133}_{\phantom{k} 55}\text{Cs})\approx 65 (M_Z / M_{Z^{\prime}_{\chi} })^2$.  
The measurement of $\QW$ given in 
Eqn.~\ref{result} implies $M_{Z^{\prime}_{\chi} } \gtrsim 1.3$ TeV at 84\% CL. 
Analogous mass bounds for various other 
$Z^\prime$ models are given in  Table~\ref{table:tag}. 

We note that these mass bounds are derived using only the $\QW$ measurement
and a global fit including the precision EW data, etc. would improve these 
constraints \cite{Erler:2009jh,arXiv:1005.3998}.
However, by considering the $Q_W(^{133}_{\phantom{k} 55}\text{Cs})$ 
measurement on its own, we can compare its sensitivity 
to that of other $Z'$ limits.

\begingroup
\squeezetable
\begin{table*}[htb]
\caption{
Mass bounds from various APV observables. The second column contain the 
mass bounds from the actual measurement in  Eqn.~\ref{result} at 95\% CL. 
The remaining columns contain 
the masses that future APV experiments will be able to exclude, given a measurement that is in agreement 
with the SM prediction. All mass bounds are the expected 95\% C.L. values.}
\begin{ruledtabular}
\begin{tabular}{ c  c c c  c  c  c  c  c}
Model & $\QW$  &$ Q_W(^{208}_{\phantom{k} 87} \text{Fr})$  &
${\cal R}_{\text{Fr}}(121,122)$  & ${\cal R}_{\text{Fr}}$(121,122) &
${\cal R}_{\text{Yb}}$(98,100) & ${\cal R}_{\text{Yb}}$(98,100) & $Q_W^p$ ({\tt Qweak}) 
				& $Q_W^p$ (P2)  \\
& 0.48\% &  0.1\%  & 0.3\% & 0.1\%  & 0.3\% &  0.1\% & 4.1\% & 2.1\% \\ \hline
 E$_6$ $\eta$   & 485    & 997   & 339   & 585   & 337   & 581  & 356  & 497 \\
 E$_6$ $\chi$   & 969   & 1993  & 679   & 1170  & 674   & 1162 & 712 & 995 \\
 E$_6$ $\psi$   & 0     & 0     & 0     & 0     & 0     & 0    &  0  & 0  \\
 E$_6$ $I$      & 1083   & 2228  & 759   & 1308  & 754   & 1299 & 796 & 1112 \\
 E$_6$ $sq$     & 1110  & 2283  & 778   & 1340  & 772   & 1331 & 815 & 1139 \\
 E$_6$ $N$      & 593     & 1220   & 416   & 716   & 413   & 712  & 436 & 609 \\
 Left Right (LR)     & 1033   & 2117  & 0     & 0     & 0     & 0   &  352 & 492  \\
 Alternate LR (ALR)  & 741   & 1527   & 701   & 1210  & 696   & 1202 &  772 & 1079 \\
 UUM            & 505   & 1012   & 953   & 1651 & 946   & 1640 &  1124 & 1570 \\
 SSM            & 1033  & 2117  & 0     & 0     & 0     & 0   &  352 & 492 \\
 TC1       & 520   & 1073   & 552   & 954   & 549   & 948 &  616 & 861 \\
 Littlest Higgs (LH) & 505    & 1012   & 953   & 1651  & 946   & 1640 &  1124 & 1570 \\
 Simplest LH (SLH)   & 1589  & 3274  & 1409  & 2433  & 1400  & 2417 &  1541 & 2153 \\
 Anom. Free LH (AFLH) & 1320  & 2718  & 1051   & 1812  & 1043  & 1800 &  1130 & 1579 \\
 331 2U1D       & 968   & 1993  & 770   & 1329  & 765   & 1320 &  829 & 1158 \\
 331 1U2D       & 1589  & 3274  & 1409  & 2433  & 1400  & 2417 & 1541 & 2153 \\
 ETC            & 245   & 490   & 461   & 800   & 458   & 794  & 544  & 760 \\
 TC2            & 872   & 1800  & 926   & 1601  & 920   & 1590 & 1034 & 1445 \\
\end{tabular}
\end{ruledtabular}
\label{table:tag}
\end{table*}
\endgroup

The simplest and anomaly free LH $Z^{\prime}$s are particularly well constrained 
since they predict large, positive values for both $\Delta Q^{p}_W$ and $\Delta Q^{n}_W$, 
which combine to give a 
large overall value for ${}^A_Z \Delta Q_W$. The opposite is true for the UUM, Sequential 
SM, Littlest Higgs and models with an SU$(2) \times $SU$(2)$ group structure such as LR and Extended TC. 
For these models $\Delta Q_W^p$ and $\Delta Q_W^n$ have opposite sign and the partial cancellation 
results in weak constraints from $\QW$. Finally, APV experiments cannot constrain the E$_6$ $\psi$ 
model, since it has $\tilde{c}_V^{u,d} = 0$ which implies no correction to any nuclear weak charge.

\subsection{Future APV Measurements}

Improving the determination of $\QW$ would require theoretical improvements in 
addition to the experimental efforts, 
and the mass bounds in the second column of Table~\ref{table:tag} are not likely 
to be improved in the near future. However, the next generation of APV experiments is underway, 
in Ba \cite{barium}, Fr \cite{francium,triumf}, Ra\cite{radium} and Yb \cite{ytterbium}, 
each of which is well-suited to APV experiments because they are expected to exhibit large parity 
violation, and there are multiple stable isotopes of each of these elements. 
We mention for completeness a proposal to measure the weak charges of
an isotope chain of Cs \cite{lanl1302} at the level of 0.2\%. 
With multiple isotopes, 
a ratio can be exploited to largely cancel the required atomic and nuclear theory input
\cite{Porsev:2010de,Brown:2008ib,Derevianko:2001uq}. 
Using measurements of weak charges along isotope chains, the following ratios can be
defined 
\cite{RamseyMusolf:1999qk}
\begin{eqnarray} \label{RR}
{\cal R}_{\text{X}}(N,N^{\prime}) = \frac{Q_W^{N^{\prime}} - Q_W^N}{Q_W^{N^{\prime}} + Q_W^N } & \text{or} & {\cal R}^{\prime}_{\text{X}}(N,N^{\prime}) = \frac{Q_W^{N^{\prime}} }{Q_W^N}
\end{eqnarray}
where $Q_W^{N (N^{\prime})}$ are the weak charges of two isotopes of element X. By circumventing large 
atomic theory uncertainty in this way, future APV experiments hope to probe the standard model at the 
sub-1$\%$ level.

Of the two observables in  Eqn.~\ref{RR}, we find that ${\cal R}_{\text{X}}(N,N^{\prime})$
is the more
sensitive probe of $Z^{\prime}$ physics.  We therefore restrict
our discussion to ${\cal R}_{\text{X}}(N,N^{\prime})$
and suggest that future experiments measure 
this quantity. We consider the corrections arising from new physics, which can
be represented by
\begin{eqnarray} \label{deltaR} \nonumber
\delta_{\cal{R}_{\text X}}^{(N,N^\prime)} &\equiv& \frac{{\cal R}_{\text X}(N,N^\prime) - {\cal R}_{\text X}^{0}(N,N^\prime) }{ {\cal R}_{\text X}^{0}(N,N^\prime)} \\ \nonumber
&=& \frac{2Z \left[ \Delta Q_W^n (1 - 4 s_W^2) + \Delta Q_W^p\right]}
{ (N^\prime + N)(1 - \Delta Q_W^n ) - 2Z(1 - 4 s_W^2 + \Delta Q_W^p) } \\
&\approx& \left( \frac{2Z}{N^{\prime} + N} \right) \Delta Q_W^p
\end{eqnarray}
where ${\cal R}_{\text{X}}^0(N,N^\prime)$ is the standard model prediction.  The approximation 
follows by setting $1 - 4 s_W^2 \approx 0$ and $\Delta Q_W^{p,n} \ll 1$, and agrees with the
result of Ramsey-Musolf \cite{RamseyMusolf:1999qk}.  We use the exact 
formula to obtain our numerical results.

The $M_{Z^{\prime}}$ dependence of  Eqn.~\ref{deltaR} 
resides in the $\Delta Q_W^{p,n}$ terms and, as with ${}^A_Z Q_W$, a determination of 
${\cal R}_{\text{X}}(N,N^{\prime})$
can be translated into a mass bound on $M_{Z^\prime}$. Since these experiments have not yet 
taken place, we derive expected mass bounds by assuming that a given experiment has made a 
measurement in agreement with the SM, with an error as given in Table~\ref{table:tag}.

There is a subtlety when calculating mass bounds from $\delta_{\cal R}$. Some models 
counterintuitively predict a small value of $\delta_{\cal R}$ for small 
$M_{Z^{\prime}}$, so that a measurement of ${\cal R}$ only excludes a mass region 
$M_{\text{min}} < M_{Z^{\prime}} < M_{\text{max}}$. However, in every model we consider, 
${M}_{\text{min}}$ is small enough that it is already ruled out by other experiments
and this issue can be safely ignored.

The mass bounds from  ${\cal R}_{\text{X}}(N,N^{\prime})$ are largest for the lightest isotopes
as can be seen from  Eqn.~\ref{deltaR}.  The values given in Table~\ref{table:tag} are therefore
calculated using the lighter isotopes to be studied for a given element.
Furthermore, the mass bounds are not sensitive to small differences in proton number $Z$, which is 
apparent from the Fr and Yb mass bounds in Table~\ref{table:tag}, so we do not also list those of Ba
and Cs. 
Likewise, the stable isotopes of Ra and Fr have nearly identical atomic numbers and we find that
the mass bounds from the two nuclei are very similar, so we only list only those of Fr.
Thus, we conclude that each of the future APV experiments is sensitive to 
the same region of parameter space and claim, very generally, 
that these experiments should aim to measure ${\cal R}$ with at least $\sim$ 0.3\% precision 
to probe new physics. We see in Table~\ref{table:tag} that at this precision, 
future APV experiments, for some models, will begin to probe a higher mass region than the current
measurement of $\QW$.  

Since the observables in Eqn.~\ref{RR} consist of weak charge ratios, there could be new physics 
models that might mimic the SM model prediction for $\cal R$ and thus remain unconstrained. 
For example, this happens in new physics scenarios that contribute to the nuclear weak charge in 
proportion to the SM prediction $\Delta Q_W^N \propto (Q_W^{N})^0$.
From Eqn.~\ref{deltaR} we can see that $\delta_{\cal R} = 0$ if
$\Delta Q_W^p = -(1 - 4 s_W^2 ) \Delta Q_W^n$. (This relation is lost if the approximate formula 
in Eqn.~\ref{deltaR} is used.) 
This defines a line in $Z^{\prime}$ 
coupling space
\begin{equation} \label{noLine}
 \tilde{c}_V^u = \frac{-3 + 8 s_W^2}{3 - 4 s_W^2} \tilde{c}_V^d \approx -\frac{1}{2} \tilde{c}_V^d
\end{equation}
describing theories that are unconstrained by measurements of ${\cal R}$.
The LR model 
and the Sequential SM fall on this line, hence their trivial mass bounds in the 
$4^{th}$ to $7^{th}$ columns of Table~\ref{table:tag}. This behaviour is obvious for the 
Sequential SM $Z^\prime$, since it has couplings identical to the SM $Z$. However, it happens through a cancellation in the LR model, which has
$\tilde{c}_A^f=-\beta c_A^f$ and $\tilde{c}_V^f=c_V^f/\beta$ where $\beta=\sqrt{1-2s_W^2}$. In this case, the product which governs the weak charge corrections
$\tilde{c}_A^e \tilde{c}_V^{u,d}=-c_A^e c_V^{u,d}$ 
is indeed proportional to the SM prediction, hence the trivial mass bounds.

Consequently, we suggest that a measurement of ${\cal R}_X$ in a given element be 
accompanied by an extraction of $Q_W(^{A}_Z X)$. The two quantities, $Q_W(^{A}_{Z} X)$ and ${\cal R}_X$, are complementary, 
since $Q_W(^{A}_{Z} X)$ is sensitive to both $\Delta Q_W^{n}$ and $\Delta Q_W^{p}$, 
whereas ${\cal R}_X$ is predominantly sensitive to $\Delta Q_W^p$, to 
the extent that the approximations in Eqn.~\ref{deltaR} are valid. This suggestion assumes that atomic and nuclear theory uncertainties
are not overwhelming \cite{Porsev:2010de,Brown:2008ib,Derevianko:2001uq}, despite the fact that the rationale for using ratios of isotopes was to reduce the impact of atomic and nuclear theory uncertainties in the extraction $Q_W$ from the APV observables, because these uncertainties are large.
Nonetheless, let us optimistically consider the case in which 
the theoretical uncertainties could be reduced to the same level of precision as
the experimental measurements so that $Q_W$ could be determined from APV measurements
to a combined uncertaintly of, for example,  0.1\%.
The mass bounds that could be obtained from a 
measurement of $Q_W(^{208}_{\phantom{k} 87}\text{Fr})$ at this precision are included in Table~\ref{table:tag}.
We see that one could approximately double the $M_{Z'}$ bounds obtained
from $^{133}\text{Cs}$.
One would obtain comparable results for the other 
isotopes being studied. We emphasize that the key to these results is reducing the theoretical
uncertainty. 

If this reduction in theoretical uncertainty is possible, then it also becomes interesting to consider a combination of $Q_W$  from
different isotopes that would cancel the proton contribution, thereby isolating the neutron
contribution.  We define a quantity
\begin{equation}
D_X(N,N')=Q_W^{N'}-Q_W^N
\end{equation}
from which the correction arising from new physics is given by
\begin{equation}
\delta_{D_X}^{N,N'}= {{D_X(N,N')- D_X^0(N,N')}\over{D_X^0(N,N')}}
=-\Delta Q_W^n.
\end{equation}
Note that $\delta_D$ is independent of $N$ and $N'$.
$D_X$ would constrain the $Z'$ couplings in a manner that would complement the other weak charge
constraints as we will show below.

\subsection{Proton Weak Charge}

In addition to the APV observables, we have included the bounds 
that can be extracted from measurements of the proton weak charge $Q_W^p$ by the 
{\tt Qweak} experiment at Jefferson Lab \cite{Gericke:2010zz}
which has recently started taking data. The new physics 
correction to the weak charge of the proton is given by
\begin{equation}
\delta_{p}\equiv \frac{\Delta Q_W^p}{(Q_W^p)^0} =-4\frac{M_Z^2}{M_{Z^{\prime}}^2} 
\frac{g_{Z^{\prime}}^2}{g_Z^2}\frac{\tilde{c}_A^e(2\tilde{c}_V^u+\tilde{c}_V^d)}{1-4s_W^2}
\end{equation} 
and we see that this observable is sensitive to the same physics as the APV observables.
In the $8^{th}$ column 
of Table~\ref{table:tag} we list expected bounds from {\tt Qweak} that assume a measurement 
of $Q_W^p$ in agreement with the SM prediction at $4.1\%$ precision. 
These expected bounds are generally comparable to the actual bounds obtained from $\QW$ 
although in a few cases [UUM, LH, ETC, TC2]
they surpass the $\QW$ bounds and will not be exceeded by APV measurements until ${\cal R}$
is measured at the highest precision in the future. 
However, the P2 experiment at the MAMI facility in Mainz is 
under development with the goal of measuring 
the proton weak charge $Q_W^p$ to 2.1\% precision \cite{p2@mami}.  The bounds that 
could be extracted from measurement of $Q_W^p$ by the P2 experiment are also included in 
Table~\ref{table:tag}.

\subsection{Other Future Parity Violating Experiments}

In addition to the measurements described above there are two additional experiments under construction
at the Jefferson Lab.  The Moller experiment is a high precision measurement of parity violation
in $e^-e^-$ scattering \cite{moller}.   The Moller collaboration estimates that they can measure the 
combination of electron couplings $c_A^e c_V^e$ to 7\%.  
We do not include bounds that could be obtained on $Z'$ masses from the Moller experiment 
in Table~\ref{table:tag} 
as for most, although not all models, they fall below the bounds obtained 
by the $\QW$ measurement.
However it is important to note, as pointed out 
by Li, Petriello and Quakenbush, that the Moller
experiment could provide important information on determining $Z'$ couplings, complementary to LHC
measurements \cite{Li:2009xh,Chang:2009yw}.  
We do not include these in our analysis on couplings because we are 
focusing on constraints on quark couplings to $Z'$'s.

The SoLID experiment measures the left-right asymmetry obtained from
deep inelastic scattering of longitudinally polarized electrons on a deuterium target
\cite{Souder:2011zz,solid,Mantry:2010ki}.  The
SoLID collaboration extimates that they will be able to measure the combination of couplings
\begin{equation}
2C_{1u}-C_{1d} \propto 2 c_A^e c_V^u-c_A^e c_V^d
\end{equation}
to a precision of, at best, 0.6\%.  The bounds that can be obtained from this level of precision 
are generally lower
than other measurements that we have considered so we do not include them in Table~\ref{table:tag}.
We mention an interesting exception, that of a leptophobic $Z'$, that can arise, in for example,
$E_6$ scenarios \cite{Buckley:2012tc}.  The leptophobic $Z'$ can contribute to the SoLID
asymmetry through photon-$Z'$ mixing.
In any case, the $u$- and $d$-quark couplings appear in a different linear combination than appears
elsewhere so the SoLID measurement can potentially be useful for constraining the couplings of a
$Z'$.  This will be explored in next section. 

These measurements are discussed in more detail in Ref.~\cite{RamseyMusolf:1999qk}.

\subsection{Comparison with Direct Detection}

We can compare this section's mass bounds to direct search limits obtained
by the LHC experiments. 
The ATLAS collaboration has obtained $Z'$ mass bounds 
based on dilepton resonance searches in $\mu^+\mu^-$ and $e^+e^-$ final states
for the  $\sqrt{s}=7$~TeV run with 5.0~fb$^{-1}$ and 4.9~fb$^{-1}$ integrated luminosity for the two 
final states \cite{ATLAS:ICHEP}.  They find:
$M(Z^\prime_{SSM})>2.21$~TeV, $M(Z^\prime_\eta)>1.84$~TeV, 
$M(Z^\prime_\chi)>1.96$~TeV and $M(Z^\prime_\psi)> 1.76$~TeV.  The CMS collaboration
has presented some limits which include results for $\sqrt{s}=8$~TeV with 3.6~fb$^{-1}$ together
with $\sqrt{s}=7$~TeV \cite{CMS:ICHEP}.  
They find $M(Z^\prime_{SSM})>2.59$~TeV and $M(Z^\prime_\psi)> 2.26$~TeV.
The LHC limits clearly
exceed those obtained from $\QW$. However, future APV experiments could be sensitive to larger
$Z^\prime$ masses 
in, for example, the Simplest LH, Anomaly Free LH and 
331 (1U2D) models, and could remain competitive with direct LHC searches until
the LHC reaches its design energy and luminosity \cite{Diener:2010sy}.

\begin{figure*}[t]
\centering
$\hspace*{-0.6cm}
\begin{array}{cc}
\includegraphics[scale=0.40,keepaspectratio=true]{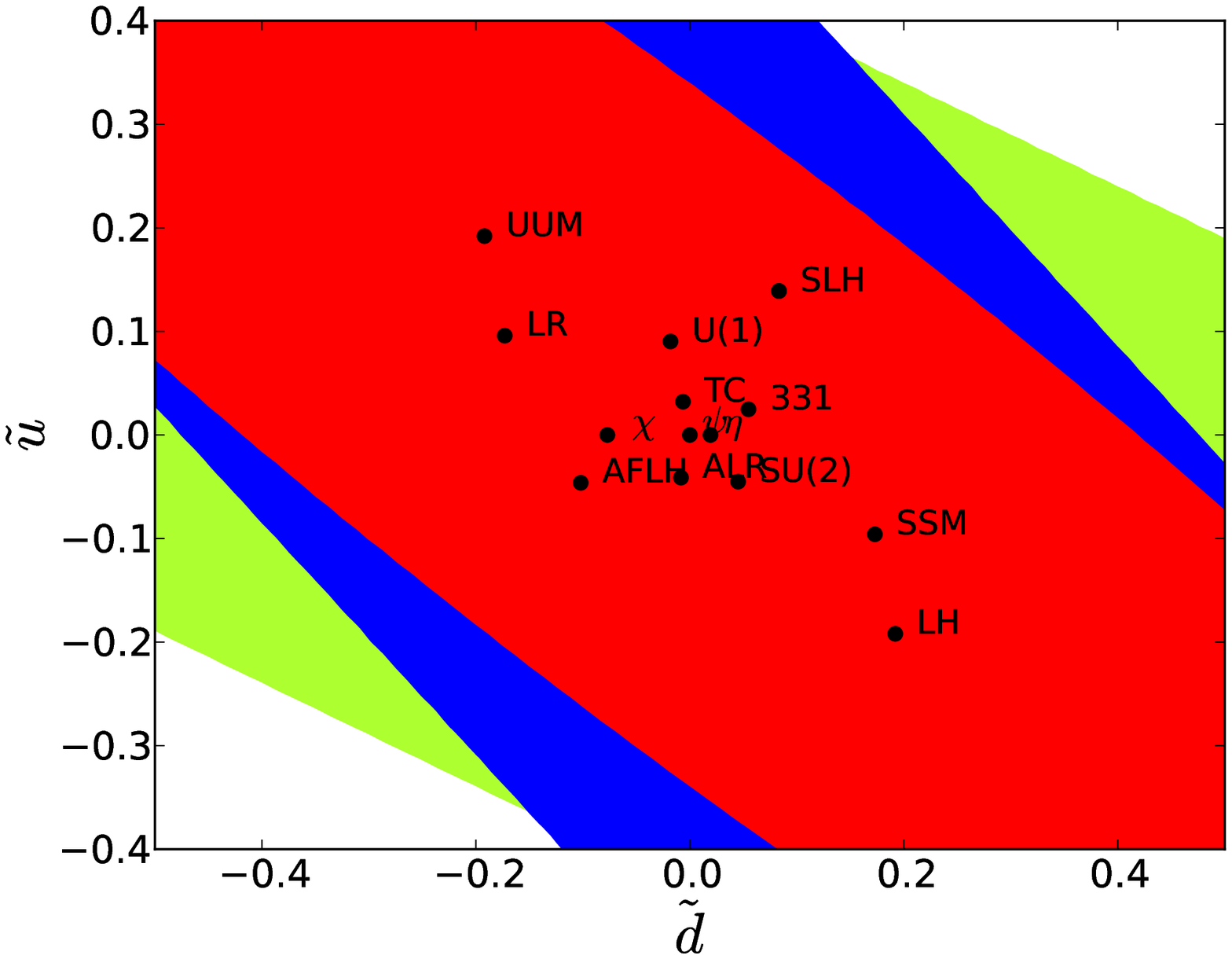} &\hspace*{-0.2cm}
\includegraphics[scale=0.40,keepaspectratio=true]{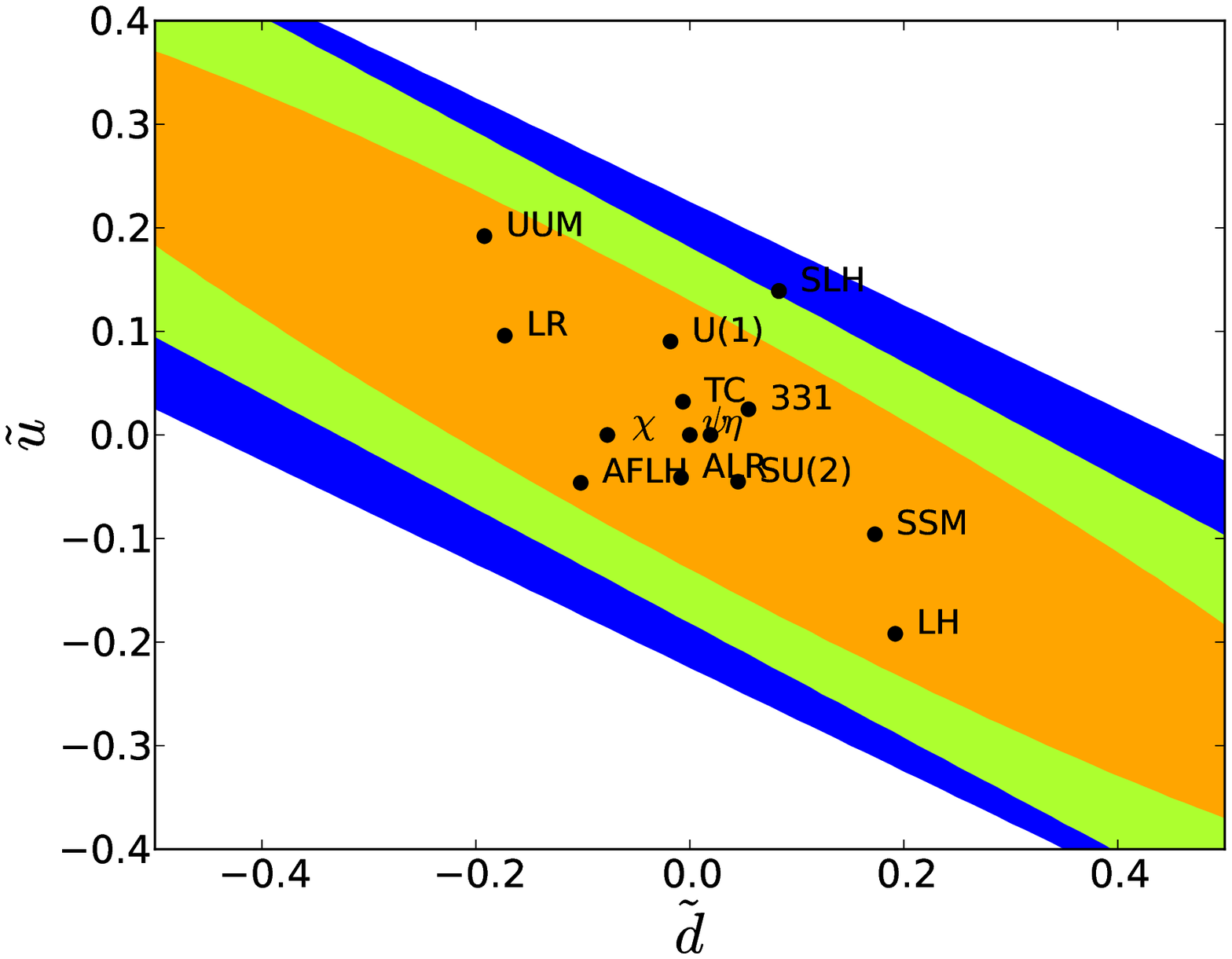} 
\\
\includegraphics[scale=0.40,keepaspectratio=true]{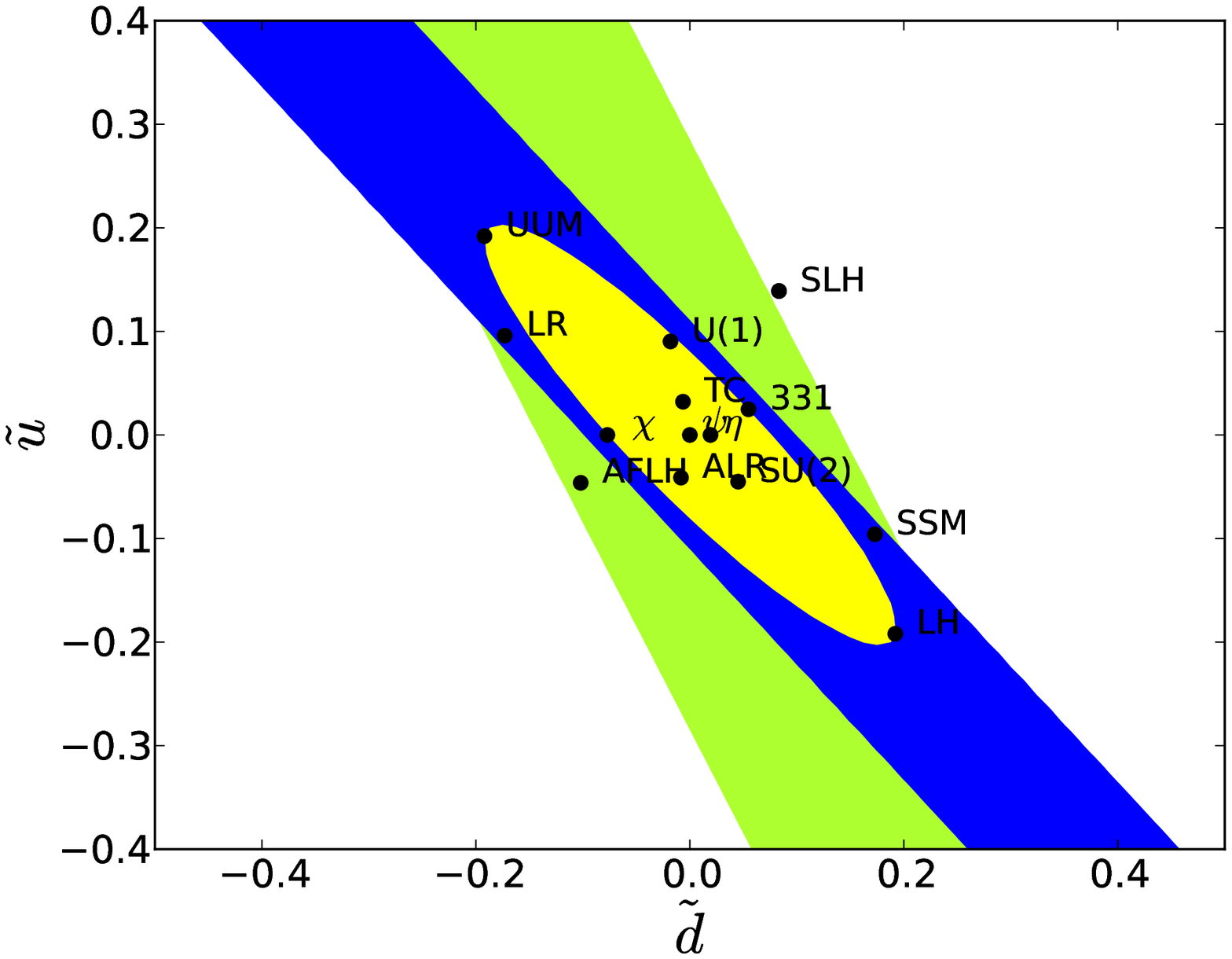} &\hspace*{-0.2cm}
\includegraphics[scale=0.40,keepaspectratio=true]{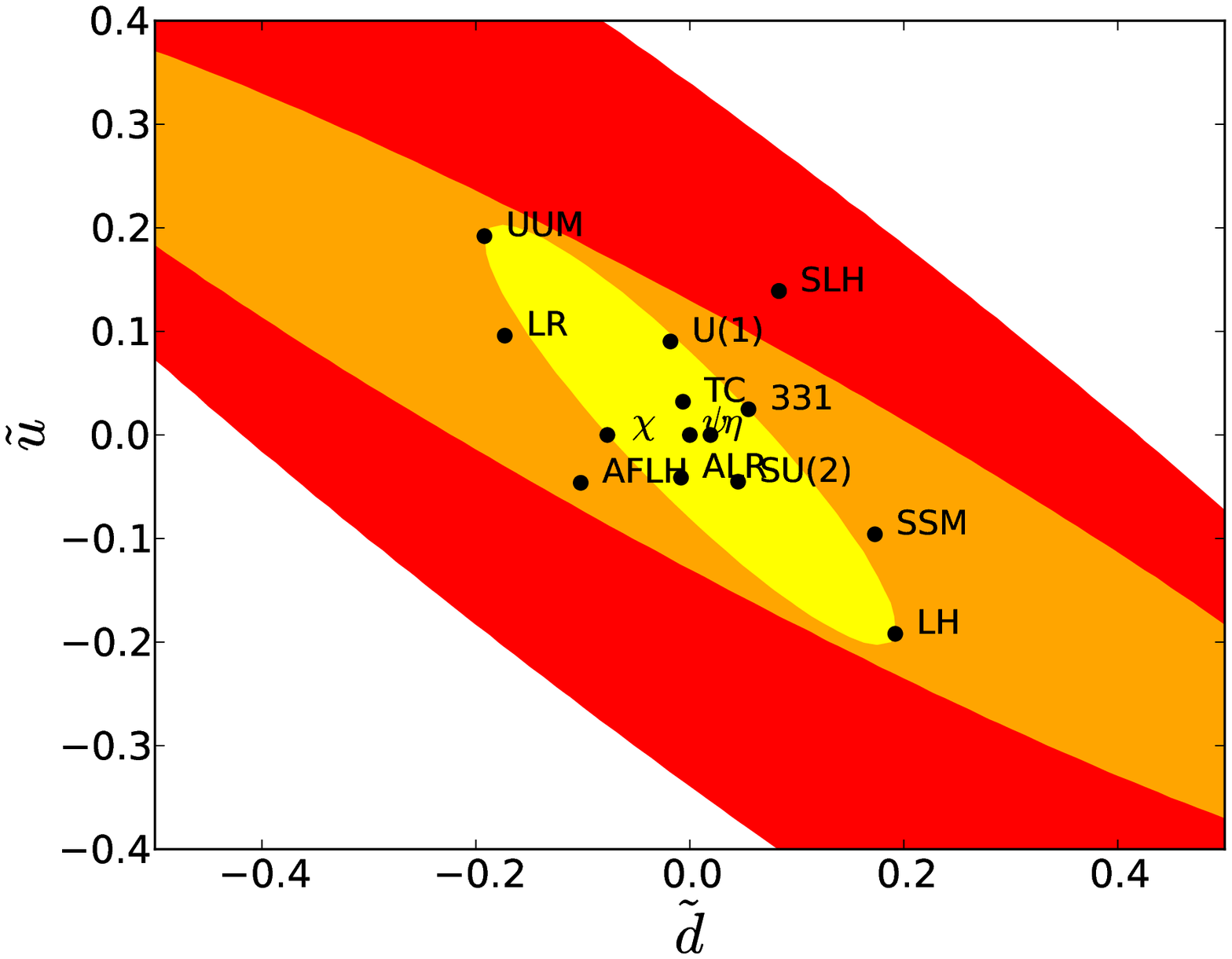}
\end{array}
$
\vskip -0.1in
   \caption{
Allowed 68\% C.L. regions for the couplings $\tilde{d}$ and $\tilde{u}$ 
for a 2.5~TeV $Z^\prime$.  It is assumed that all measurements are in agreement with the SM.
In (a) the blue (dark grey) band corresponds to the region allowed by 
the $^{133}$Cs weak charge measurement with 0.48\% precision, 
the green (light grey) band corresponds
to the allowed regions expected from the {\tt Qweak} measurement of $Q_W^p$ with 4.1\% precision,
and the red (medium grey) oval is the region allowed by a combined fit of $Q_W(^{133}\text{Cs})$ and
$Q_W^p$.
In (b) the blue (dark grey) region is the region 
that would be constrained by the P2 measurement of $Q_W^p$ with 2.1\% precision, 
the green (light grey) band is the region allowed by 
${\cal R}_{\text{Fr}}(121,122)$ measurement with 0.1\% precision, 
and the orange (medium grey) oval is the region allowed
by a combined fit using the SoLID results, $Q_W^p$ from P2 and 
{\tt Qweak}, ${\cal R}_{\text{Fr}}(121,122)$, and $Q_W(^{133}\text{Cs})$.
In (c) the blue (dark grey) band is the region that would be constrained by 
a $^{208}$Fr weak charge measured to the precision of 0.1\%, the  green (medium grey) 
band is the region
that would be constrained by  a $D_{\text{Fr}}$ measured to 0.1\%, and the yellow (lightest grey) oval
is the region that would constrained from a combined fit to $Q_W(^{208}\text{Fr})$, $D_{\text{Fr}}$,
and the other five measurements. 
(d) collects the combined fits from (a), (b), and (c) to show how 
successive measurements can improve the contraints on $Z'$ couplings.
}
\label{coupling}
\end{figure*}

\section{Bounds on the Couplings}\label{sec:coupl}

If a $Z^{\prime}$ boson with mass $M_{Z^\prime}$ were discovered at the LHC, 
weak charge measurements could be used to constrain its couplings.
Since the mass would be fixed, APV and {\tt Qweak}
experiments would constrain the coupling combinations 
$(\tilde{d}, \tilde{u}) \equiv (g_{Z^{\prime}}^2/g_Z^2) \tilde{c}_A^e \tilde{c}_V^{(d,u)}$. 
For example, the $\QW$ measurement constrains at 68\% C.L.
the $\tilde{u}$ and $\tilde{d}$ couplings of 
a 2.5~TeV $Z^\prime$ to lie within a band in parameter space, as shown in Fig.~\ref{coupling}(a).

We also include in Fig.~\ref{coupling}(a) 
the constraints one could obtain from the expected {\tt Qweak} measurement of 
$Q_W^p$. We assume that $Q_W^p$ is measured to 
have its SM value with the stated experimental error, 
and plot the expected 68\% C.L. bounds on $\tilde{u}$ 
and $\tilde{d}$ for a 2.5~TeV $Z^\prime$.  Additionally, we show the region of the $\tilde{u}-\tilde{d}$
parameter space that would be constrained at 68\% C.L.
by combining the APV measurement of $\QW$ and the  {\tt Qweak} of $Q_W^p$. We obtain
this contour by calculating the $\chi^2$ obtained by comparing the SM values to a scan
of the  $\tilde{u}-\tilde{d}$ parameter space. 
In addition, we
plot the predicted value of $\tilde{u}$ and $\tilde{d}$ for each of the models we consider.
We see that these measurements would not in themselves be able to identify a 2.5~TeV $Z^\prime$.

We next consider how approved future experiments will constrain the $Z^\prime$ couplings which
we show in Fig.~\ref{coupling}(b).  We include constraints one could obtain from a measurement of 
${\cal R}_{\text{X}}(N,N^{\prime})$ using isotopes of Fr as a representative example, the 
expected P2 measurement of $Q_W^p$, and the measurement of the coupling combination
$2C_{1u}-C_{1d}$ by the SoLID experiment.  
As before we assume that ${\cal R}_{\text{Fr}}$, $Q_W^p$, and $2C_{1u}-C_{1d}$ 
are found to have their SM values with the stated experimental error, 
and plot the expected 68\% C.L. bounds on $\tilde{u}$ 
and $\tilde{d}$ for a 2.5~TeV $Z^\prime$. 
We also calculated the expected constraints 
based on $\cal R_{\text{Yb}}$ but they are almost identical to those of Fr so we do not show them 
in the figure.  The final contour in  Fig.~\ref{coupling}(b) shows the 68\% C.L. region found
by combining the expected experimental precision for all five of these measurements;
$Q_W$ of $^{133}_{55}Cs$, $Q_W^p$ from {\tt Qweak} and P2, and the couplings from SoLID.
We do not show the constraints derived from
measurement by the SoLID experiment as they fall outside the range of the figure but
do include them in the combined fit as they do improve the constraints slightly.

Finally, in Fig.~\ref{coupling}(c) we 
show the bounds on $\tilde{d}$ and $\tilde{u}$ from $Q_W(^{208}_{\phantom{k} 87}\text{Fr})$   
and $D_{\text{Fr}}(N,N')$, in both cases assuming a hypothetical 0.1\% combined theoretical and
experimental uncertainty.  We also show the 68\% C.L. region found
by combining the expected experimental precision for these two measurements plus the five already
described above. 

In  Fig.~\ref{coupling}(d) we show the 68\% C.L. regions for the three cases
described above; (i) the APV measurement of $\QW$ and the  {\tt Qweak} measurement of $Q_W^p$. 
(ii) These
two measurements plus ${\cal R}_{\text{Fr}}$, $Q_W^p$ from P2, and the constraints from
the SoLID experiment.
(iii) All of these plus $Q_W(^{208}_{\phantom{k} 87}\text{Fr})$  and $D_{\text{Fr}}(N,N')$.
One can see how successive improvements in the experimental and theoretical
uncertainties can improve the constraints on $Z^\prime$ couplings if one were to be discovered.
Theoretical uncertainties are the dominant uncertainties in  $Q_W(^{208}_{\phantom{k} 87}\text{Fr})$   
and $D_{\text{Fr}}(N,N')$.  Thus,  reducing the theoretical uncertainties needed
to obtain $Q_W(^{208}_{\phantom{k} 87}\text{Fr})$ from the APV measurement can result in a
significant improvement in determining the $Z^\prime$ couplings.
We conclude that when APV measurements of $\QW$ and future 
measurements of ${\cal R}$ and $Q_W^p$ are combined with measurements from the LHC and other low 
energy precision experiments \cite{Li:2009xh}, they could add useful information about a $Z^\prime$ 
boson's $u$ and $d$ couplings that are not easily obtained elsewhere.

Finally, it is worth pointing out the generality of the results in Fig.~\ref{coupling}.
While we have focussed on $Z^\prime$ physics,
the combinations $g_Z^2 \tilde{d} $ and $g_Z^2 \tilde{u}$ are just the overall dimensionless 
couplings that appear in the new physics effective Lagrangian of Eqn.~\ref{effLag}. 
We explicitly include the factor of $g_Z^2$ because we normalized $\tilde{u},\tilde{d}$ to the 
SM $Z$ coupling strength. 
$M_{Z^{\prime}}$ is just the overall 
mass scale, so the constraints in Fig.~\ref{coupling} can be immediately recast into constraints on 
other parity violating new physics scenarios. If some other new physics described by
the effective Lagrangian
\begin{equation}
\Delta {\cal L}_{\text{NP}}^f = -\frac{(g_Z^2 \tilde{f})}{4 \Lambda^2} ( \bar{e}
\gamma_\mu \gamma_5 e )(  \bar{f} \gamma^\mu f )
\end{equation}
were discovered at a mass scale $\Lambda = 2.5$~TeV, then the normalized couplings 
$\tilde{f}$ in the $u$ and $d$ sector would also be constrained exactly as shown in 
Fig.~\ref{coupling}. A $Z^{\prime}$ boson is a particular case, with 
$\Lambda = M_{Z^{\prime}}$ and 
$\tilde{f}= (g_{Z^{\prime}}^2/g_Z^2) \tilde{c}_A^e \tilde{c}_V^{(d,u)}$.

\section{Conclusion}\label{sec:conc}

Extra neutral gauge bosons, $Z^\prime$s,  arise in many extensions of the standard model.  
In this paper 
we explored the constraints that atomic parity violation experiments can place on $Z^\prime$s.  While
this subject has been studied previously, we consider a large collection of models, and we explore
a number of new experiments which plan to observe APV in different isotopes of Ba, Fr, Ra and Yb.  
These new experiments allow the measurement of weak charge ratios along isotope chains
which will abate difficulties associated with atomic theory uncertainties.

We have two main results:
the constraints that APV measurements can put on a $Z^\prime$ mass for a given model, and the 
constraints
that APV measurements can put on $Z^\prime$ couplings if one were to be discovered at the LHC.  
We also include the bounds expected from the {\tt Qweak} measurement of the proton weak charge
and futre measurements by the P2 and SoLID experiments. 
We find that the current $0.48\%$ precision measurement of $\QW$ constrains $Z^\prime$ masses
to be above $\sim$500 GeV to $\sim$1600 GeV at $95\%\,{\rm C.L.}$, depending on the model.  
Future APV experiments which will measure isotope ratios for Fr and Yb could 
yield bounds close to $\sim 2$~TeV for the UUM, LH, SLH, and AFLH models.  While bounds from the 
LHC's direct searches already exceed the $\QW$ limits on most models, we found that future APV 
limits could still be competitive for some models, such as variations of the Little Higgs
models \cite{Diener:2010sy}.

We also considered the constraints that APV experiments could put on $Z^\prime$ couplings if 
a $Z^\prime$ were discovered, which will be an important step in better understanding the 
underlying physics. We found that measurements at {\tt Qweak} and the APV experiments could be 
used to distinguish between some, but not all models. Future measurements by the P2 and SoLID
experiments will also provide useful input.  The addition of 
$Q_W(^A_{\phantom{k} 87}\text{Fr})$ and $D_{\text{Fr}}$ at 0.1\% precision would result
in better discrimination, highlighting the importance of improving both theoretical and experimental
uncertainties.   Regardless, current and future APV experiments will provide 
information complementary to other measurements \cite{Li:2009xh} and would 
be an important addition to fits used to constrain a newly discovered $Z^\prime$.

\acknowledgments

This research was supported in part by the Natural Sciences and Engineering Research Council of Canada. 
The authors thank John Behr, Alain Bellerive and Gerald Gwinner for helpful communications and 
RD would like to thank Andrei Derevianko for his efficient correspondence. 
Research at the Perimeter Institute is supported by the Government of Canada and by the 
Province of Ontario through the Ministry of Research and Innovation.

\end{document}